\RequirePackage{fix-cm}
\documentclass[smallextended]{svjour3}       
\smartqed  
\usepackage{graphicx}
\usepackage{amssymb}
\usepackage{setspace}
\usepackage{amsmath}
\newtheorem{thm}{Theorem}
\newtheorem{prop}[thm]{\textsc{Proposition}}
\newtheorem{lem}[thm]{\textsc{Lemma}}
\newtheorem{cor}[thm]{\textsc{Corollary}}
\newtheorem{exam}[thm]{\textsc{Example}}
\newtheorem{rem}[thm]{\textsc{Remark}}

\newcommand{\pf}{\noindent {\bf Proof.  }}

\def\0{{\mathbf 0}}
\def\C{{\mathcal C}}

\newcommand{\Z}{\mathbb{Z}}

\newcommand{\CC}{\mathbb{C}}

\let\MakeUppercase\relax

\begin{document}
\title{Cyclic and Quasi-Cyclic DNA Codes
}


\author{
Adel Alahmadi         \and
        Alaa Altassan \and
        Amani Alyoubi\and
        Manish K. Gupta   \and
        Hatoon Shoaib
}


\institute{A. Alahmadi \at
              Saudi Arabia \\
              \email{adelnife2@yahoo.com}           
           \and
           A. Altassan \at
              Saudi Arabia\\
           \email{aaltassan@kau.edu.sa} 
           \and 
           A. Alyoubi \at
           Saudi Arabia\\
           \email{moon61343@gmail.com}
            \and 
           M. Gupta \at
           India\\
           \email{mankg@computer.org}
            \and 
           H. Shoaib \at
           Saudi Arabia\\
           \email{hashoaib@kau.edu.sa}
}

\date{Received: date / Accepted: date}

\maketitle

\begin{abstract}
In this paper, 
we discuss DNA codes that are cyclic or quasi-cyclic over  $\Z_{4}+\omega \Z_{4}$, where $\omega^{2}=2+2\omega$ along with methods to construct these with combinatorial constraints. We also generalize results obtained for the ring $\Z_{4}+\omega \Z_{4}$, where $\omega^{2}=2+2\omega$, and some other rings to the sixteen rings $R_{\theta}=\Z_{4}+\omega \Z_{4}$, where $\omega^{2}=\theta\in \Z_{4}+\omega \Z_{4}$, using the generalized Gau map and Gau distance in \cite{3}. We determine a relationship between the Gau distance and Hamming distance for linear codes over the sixteen rings $R_{\theta}$ which enables us to attain an upper boundary for the Gau distance of free codes that are self-dual over the rings $R_{\theta}$. 
\end{abstract}

\section{Introduction}
\label{intro}
The interdisciplinary field of DNA computing has developed into a captivating subject of research during the last twenty-six years. The first successful ingress into DNA computing was performed by Adleman in 1994. DNA is a double-stranded helix made up of  chains of four chemicals (nucleotides), Adenine, Cytosine, Guanine and Thymine, which we denote  by  $A$, $C$, $G$ and $T$ respectively. Each of these is paired with another  in the other strand,  its complementary base (called the Watson-Cricks complement), the two kinds of complements given by $A^{c}=T$ and $C^{c}=G$. Thus a DNA string can be represented as a finite sequence of the DNA letters ${A, C, G, T}.$ Any set of DNA strings having this set  of letters is a DNA code. For a such a string, the complement comes  from changing  each letter $X$ to  $X^c$. Similarly, the reverse string reflects it into reverse order, while the reverse complement is the combination of  both actions (in either order). A DNA code is called reversible if it contains the reverse of each of its element. It is called complement if it contains the complement of each of its element. And if the reverse complement of each element of a DNA code belongs to this code, then we call it a reversible complement DNA code. Processing in DNA computing is mostly via DNA hybridization. However, this leads to the main source of errors. To avoid these errors, DNA codes have to satisfy many constraints, especially combinatorial and thermodynamic. Over the past sixteen years many different approaches have been tried to construct these DNA strings: combinatorial, algebraic, computational. In particular, many researchers used algebraic constructions over rings and fields for the DNA codes. However, those methods often gave DNA codes with non-optimal distances. In 2018, Limbachyia et al \cite{1} discovered the Gau isometry map from the ring $R_{2+2\omega}=\Z_{4}+\omega \Z_{4}$, where $\omega^{2}=2+2\omega$, to DNA strings of length two utilizing the Gau distance on codes over $R_{2+2\omega}$ and produced several DNA codes with better parameters with some optimal ones. In 2019 \cite{3}, the Gau isometry map has been extended to cover all the sixteen rings $R_{\theta}=\Z_{4}+\omega \Z_{4} $, where $\omega^{2}=\theta \in \Z_{4}+\omega \Z_{4}$, and a generalized definition of Gau distance was given for codes over $R_{\theta}$. Cyclic DNA codes have been intensively studied for many element sets such as fields and rings. In particular in \cite{2}, cyclic DNA codes were investigated using the Gray map  over the ring $\Z_{4}+\omega \Z_{4}$ where $\omega^{2}=2$.  It is still an interesting problem to study this type of code using a different map. Besides, quasi-cyclic DNA codes have not been studied substantially. 
\par The ring $\Z_{4}+\omega \Z_{4}=\{0, 1, 2, 3, \omega, 1+\omega, 2+\omega, 3+\omega, 2\omega, 1+2\omega, 2+2\omega, 3+2\omega, 3\omega, 1+3\omega, 2+3\omega, 3+3\omega \}$ is an extension of $\Z_{4}$. Using the Gau map defined in \cite{1}, we have studied DNA codes that are cyclic or quasi-cyclic over  $\Z_{4}+\omega \Z_{4}$, where $\omega^{2}=2+2\omega$, and we have found necessary and sufficient conditions to obtain such codes satisfying the reverse, complement and reverse complement constraints. Furthermore, we have generalized many results obtained for the ring $R_{2+2\omega}=\Z_{4}+\omega \Z_{4}$\  ($\omega^{2}=2+2\omega$) and some other rings to all the sixteen rings $R_{\theta}$ using the generalized Gau map defined in \cite{3}. In addition, we have improved the maximum possible Hamming distances for self-dual codes that are free over  $\Z_{4}+\omega \Z_{4}$, where $\omega^{2}\in\{1, 2\omega \} $ in \cite{4}, to the  ring $R_{\theta}$ and we have found a relationship between the Gau and Hamming distances over the ring $R_{\theta}$, leading us to find, for  a self-dual code that is free over $R_{\theta}$, an upper limit to its Gau distance.
\par   This paper is set out as follows: Section 2 gives some examples of cyclic and 2-quasi-cyclic DNA codes over the ring $R_{2}=\Z_{4}+\omega \Z_{4}\  (\omega^{2}=2)$ using the Gau map. 
In \S3  cyclic DNA codes over the ring $R_{2+2\omega}$ are looked at, along with sufficient conditions for the construction of such codes satisfying the reverse, complement or reverse complement constraints. 
Similarly, in \S4 we find  quasi-cyclic codes over the ring $R_{2+2\omega}$ with sufficient conditions for reverse, complement and reverse complement DNA quasi-cyclic codes. 
Finally, \S5 includes the generalization of results obtained for the ring $R_{2+2\omega}$ to the general ring $R_{\theta}$, where $\theta \in \Z_{4}+\omega \Z_{4}$, and as well contains a generalization of results achieved for self-dual codes over the ring $\Z_{4}+\omega \Z_{4}$, where $\omega^{2}=2\omega$ in \cite{4}, to the rings $R_{j}=\Z_{4}+\omega \Z_{4}$, where $j \in\{2\omega, 2\}$. Also, it displays a relationship between the Hamming and Gau distances over the ring $R_{\theta}$ and upper bounds for  the Hamming and Gau distances of free self-dual codes over the ring $R_{\theta}$.

\section{DNA Codes over $R_{2}=\Z_{4}+\omega \Z_{4}\  (\omega^{2}=2)$}
\label{sec:1}

 In \cite{2} cyclic DNA codes have been studied over the ring $R_{2}$ using the Gray image. Here, we shall use the Gau map which defined in \cite{3} from the ring $R_{2}$ to the set of DNA strings of length $2$ ,where $\lambda=2$, relating it to the Gau distance to give some examples of cyclic and quasi-cyclic DNA codes.  
The zero divisors $Z$ and the units $U$ of the ring $R_{2}$ have been given in \cite{2}. 
Clearly $U = Z+1$.  From basic ring theory, $(U,.)$ is a multiplicative group, while  $(Z,+)$ is an additive group, and in this case both have order eight.  It is unusual that both these abelian groups are isomorphic to ${\Bbb Z}_4\times {\Bbb Z}_2$.  Indeed using this isomorphism we find that the multiplicative and additive orders of elements in the ring are related by the formula $|x+1|. = |x|_+$, for all $x\in Z$.

The Gau map that we will use has the following properties:
\begin{itemize}
\item[-] There are four elements that have additive order 1 or 2: $0$, $2$, $2\omega$ and $2+2\omega$. Their images $AA$, $TT$, $CC$ and $GG$ are self-reversible.
\item[-] There are four elements which have multiplicative order 1 or 2: $1$, $3$, $1+2\omega$ and $3+2\omega$. Their images $AT$, $TA$, $GC$ and $CG$ are self-reversible-complement.
\end{itemize}
\subsection{Examples}
\label{sec:2}

A cyclic code $\C$ is a linear code such that whenever $x_{0}x_{1}\dots x_{n-1}\in \C$, we have also $x_{n-1}x_{0}\dots x_{n-2}\in \C$ while the quasi-cyclic code $Q$ of index $l$ is a linear code such that for any $x_{0}x_{1}\dots x_{n-1}\in Q$, we have $x_{n-l}x_{n+1-l}\dots x_{0}\dots x_{n-1-l}\in Q$. 
We can get the 2-quasi-cyclic DNA code with parameters $(2n=8, M=32, d_{H}=4)$ by  shifting the last two coordinates of the vector $(1+3\omega, 1+\omega, 1+\omega, 1+3\omega)$. These 32 DNA codewordes are as follows:
\[ \begin{array}{lcrr}
\mbox{$AAAAAAAA$} & \mbox{$CCCCCCCC$} & \mbox{$CCGGGGCC$} & \mbox{$CTTCTCCT$}\\
\mbox{$TAGCGCTA$} & \mbox{$GCTATAGC$} & \mbox{$ATCGCGAT$} & \mbox{$ACTGTGAC$}\\
\mbox{$TTTTTTTT$} & \mbox{$GGGGGGGG$} & \mbox{$TACGCGTA$} & \mbox{$TTAAAATT$}\\
\mbox{$ATGCGCAT$} & \mbox{$CGATATCG$} & \mbox{$GCATATGC$} & \mbox{$AATTTTAA$}\\
\mbox{$CACACACA$} & \mbox{$ACACACAC$} & \mbox{$GGCCCCGG$} & \mbox{$TCCTCTTC$}\\
\mbox{$TCGAGATC$} & \mbox{$GATCTCGA$} & \mbox{$CGTATACG $} & \mbox{$GTCACAGT$}\\
\mbox{$GTGTGTGT$} & \mbox{$TGTGTGTG$} & \mbox{$GAAGAGGA$} & \mbox{$AGGAGAAG$}\\
\mbox{$AGCTCTAG$} & \mbox{$CTAGAGCT$} & \mbox{$TGACACTG$} &\mbox{$CAGTGTCA$}
\end{array}\] 
In similar ways, we have found many examples of cyclic and $2-$quasi-cyclic DNA codes given in table \ref{tab:1} and table \ref{tab:2}. For the cyclic codes, we will shift only the last coordinate of the vector.
\newpage
\begin{table}
  \caption{Cyclic DNA Codes over $R_{2}$}
\label{tab:1}       
\begin{tabular}{ll}
\hline\noalign{\smallskip}
First row vector & $(2n, M, d_{H})$  \\
\noalign{\smallskip}\hline\noalign{\smallskip}
$(2+\omega, 3, 1)$ & $(6, 2048, 1)$   \\
$(3, 1, 2+2\omega)$ & $(6, 1024, 2)$  \\
$(2+\omega, 1, 3, 2+\omega)$ & $(8, 4096, 2)$ \\
$(1, 1, 1)$ & $(6, 16, 3)$ \\
\noalign{\smallskip}\hline
\end{tabular}
\end{table}

\begin{table}
  \caption{2-Quasi-Cyclic DNA Codes over $R_{2}$}
\label{tab:2}       
\begin{tabular}{ll}
\hline\noalign{\smallskip}
First row vector & $(2n, M, d_{H})$  \\
\noalign{\smallskip}\hline\noalign{\smallskip}
$(2\omega, 1, 1, 2\omega)$ & $(8, 256, 2)$   \\
$(1+3\omega, 1+\omega, 1+\omega, 1+3\omega)$ & $(8, 32, 4)$ \\
$(3+\omega, 1, 1, 3+\omega)$ & $(8, 64, 4)$ \\
$(2, 1, \omega , \omega , 1, 2)$ & $(12, 4096, 4)$ \\
$(1+2\omega, 1, 2, 2\omega , 2\omega, 2, 1, 1+2\omega)$ & $(16, 4096, 4)$ \\
\noalign{\smallskip}\hline
\end{tabular}
\end{table}
 In both Table $1$ and Table $2$, all codes represented are reversible and reversible complement. Furthermore, Table $2$ includes DNA codes with good parameters.

\section{Cyclic Codes over $R_{2+2\omega}=\Z_{4}+\omega \Z_{4} \ (\omega^{2}=2+2\omega)$}
\label{sec:2}
In this section we consider cyclic DNA codes.

\begin{definition} Let $g(x)=g_{0}+g_{1}x+\dots +g_{n-1}x^{n-1}$ be a polynomial of degree $n-1$ with leading coefficient not equal to zero and $g_{i}\in R_{2+2\omega} $ for all $i\in \{0,1,\dots,n-1\}$. 
Then the reciprocal of $g$ denoted as $g^{*}$ is defined by $g^{*}(x)=g_{0}x^{n-1}+g_{1}x^{n-2}+\dots +g_{n-1}$.
\newline A self-reciprocal polynomial $g(x)$ is one for which  $\exists v\in R_{2+2\omega} $  such that $g^{*}=vg$.
\end{definition}

\begin{definition} Let $x=x_{0}x_{1}\dots x_{n-1} \in R^{n}_{2+2\omega}$ be any string of length $n$. The reverse of $x$ is $x^{r}=x_{n-1}x_{n-2}\dots x_{0}$. The complement of $x$ is $x^{c}=x^{c}_{0}x^{c}_{1}\dots x^{c}_{n-1}$. The reverse complement of $x$ is $x^{rc}=x^{c}_{n-1}x^{c}_{n-2}\dots x^{c}_{0}.$ Let $\CC$ be any linear code over $R_{2+2\omega}$. The code $\CC$ is called reversible if for every $x\in \CC$, we have $x^{r}\in \CC.$ The code is called complement if for every $x\in \CC$, we have $x^{c}\in \CC.$ The code $\CC$ is called reversible complement if for every $x\in \CC$, we have $x^{rc}\in \CC.$
\end{definition}

\begin{prop}\label{prop:linear}
For any linear reversible complement code $\CC$ over  $R_{2+2\omega}$, the constant string $\boldsymbol{2+2\omega}$ $\in \CC$.
\end{prop}
\pf Assume that $\CC$ is reversible complement linear code over $R_{2+2\omega}$ . Since $\CC$ is linear, then $00\dots 0\in \CC$. $\phi(00\dots 0)=AA\dots A$ and $\phi(00\dots 0)^{r}=AA\dots A=\phi(00\dots 0)$. So, $\phi^{-1}(\phi(00\dots 0)^{rc})=\phi^{-1}(\phi(00\dots 0)^{c})=2+2\omega\dots 2+2\omega= \boldsymbol{2+2\omega}\in \CC$.

\begin{thm}
For a cyclic code $\C$ over $R_{2+2\omega}[x]/<x^{n}-1>$, $\exists a(x)$, $b(x)$ such that $b(x)|a(x)|x^{n}-1$ and $\C=<a(x), \omega b(x)>$.
\end{thm}
\pf See \cite{7}  for the proof.
\vspace*{1cm}
\\ Using \cite[Lemma 4]{2}, we obtain the following:
\begin{lem}\label{lem:degree}
Let $a, b \in R_{2+2\omega}[x]$ and assuming that deg$(a(x))-deg(b(x))=t$ then:
\item[-]$(a(x)b(x))^{*}=a^{*}(x)b^{*}(x)$.
\item[-]$(a(x)+b(x))^{*}=a^{*}(x)+x^{t}b^{*}(x)$.

\end{lem}

\begin{thm}\label{thm:reversible cyclic}
If $\C=\langle a(x) , \omega b(x)\rangle$ is a cyclic code over the ring $R_{2+2\omega}$, then a necessary and sufficient condition that $\C$ is reversible is that $a$ and $b$ are self-reciprocal.
\end{thm}
\pf $(\Longrightarrow)$ Assume that $\C$ is reversible.\\* Let $a(x)=a_{0}+a_{1}x+...+a_{s}x^{s}, s \leq n-1$. Then if $a(x) \in \C$, $$ a(x)^{r}=0+0x+\dots+ox^{n-s-2}+a_{s}x^{n-s-1}+\dots+a_{0}x^{n-1}\in \C.$$ Then, $a_{s}x^{n-s-1}+\dots+a_{0}x^{n-1}\in \C$, which implies, $$x^{s+1-n}(a_{s}x^{n-s-1}+\dots+a_{0}x^{n-1})\in \C.$$ Then, $a_{s}+\dots+a_{0}x^{s} \in \C$ which means $a^{*}(x)\in \C$. Hence, $$a^{*}(x)=a(x)c(x)+\omega b(x)d(x),$$ where $c(x), d(x) \in Z_{4}[x]$. We have $a_{i}\in Z_{4}$ for $i\in \{0,1,\dots,s\}$, so $d(x)=0$. Then $a^{*}(x)=a(x)c(x)$. Hence, $deg(a^{*}(x))\geq deg(a(x))$ but $deg(a^{*}(x))\leq deg(a(x))$, and so $deg(a^{*}(x))= deg(a(x))$. Hence, $c(x)=m \in Z_{4}$ where $m\neq 0$. Thus, $a$ must be self-reciprocal.

Similarly, let $b(x)=b_{0}+b_{1}x+\dots+b_{t}x^{t} $  ,$t\leq n-1$ then, $$\omega b(x)=\omega b_{0}+\omega b_{1}x+...+\omega b_{t}x^{t}$$ $$(\omega b(x))^{r}=\omega(b(x))^{r}=0+0x+\dots+0x^{n-t-2}+\omega h_{t}x^{n-t-1}+\dots+\omega h_{0}x^{n-1} \in \C$$ Then, $$\omega (b_{t}x^{n-t-1}+\dots+ b_{0}x^{n-1}) \in \C.$$ Then, $$\omega x^{t+1-n}(b_{t}x^{n-t-1}+\dots+ b_{0}x^{n-1})\in \C,$$ which means $\omega (b_{t}+\dots+ b_{0}x^{t})\in \C$. Then, $\omega b^{*}(x)\in \C$. Hence, $$\omega b^{*}(x)=a(x)c{'}(x)+\omega b(x)d^{'}(x),$$ where $c^{'}(x) , d^{'}(x) \in Z_{4}[x]$. Since every term of $a(x)$ does not contain $\omega$, then $c{'}(x)=0$. Hence, $b^{*}(x)=b(x)d^{'}(x)$ which implies $deg(b^{*}(x))= deg(b(x))$. Then, $d^{'}(x)=m^{'} \in Z_{4}$ where $m^{'}\neq 0$. Thus, $b$ must be self-reciprocal. 

$(\Longleftarrow )$ Assume that $a(x)$ and $b(x)$ are self-reciprocal. \\*Let $h(x)=h_{0}+h_{1}x+\dots+h_{u}x^{u} \in \C   , u \leq n-1$. Then, $$h(x)=a(x)e(x)+\omega b(x)f(x),$$ where $e(x) , f(x) \in Z_{4}[x]$. \\* By lemma \ref{lem:degree}, $h^{*}(x)=(a(x)e(x)+\omega b(x)f(x))^{*}=[a(x)e(x)]^{*}+x^{t}[\omega b(x)f(x)]^{*}=a^{*}(x)e^{*}(x)+x^{t}[\omega h^{*}(x)f^{*}(x)]=a^{*}(x)e^{*}(x)+\omega b^{*}(x)(x^{t}f^{*}(x))=ma(x)+\omega(m^{'}b(x))(x^{t}f^{*}(x))$. \\*Hence, $h^{*}(x)\in \C$. So, $$h^{*}(x)=h_{u}+h_{u-1}x+\dots+h_{0}x^{u} \in \C.$$ Since $\C$ is cyclic, then $x^{n-u-1}h^{*}(x)\in \C$. Since $\C$ is linear, then $$[0+0x+\dots+0x^{n-1}]+x^{n-u-1}h^{*}(x)\in \C.$$ Therefore, $h(x)^{r}\in \C$. Thus, $\C$ must be reversible.

\begin{lem}\label{lem:5}
For any $x\in R_{2+2\omega} ; \ \omega x=(\omega x)^{c}-(2+2\omega)$.
\end{lem}
\pf It is obvious from Table I in \cite{1}.

\begin{thm}\label{thm:reverse complement}
If $\C=\langle a(x) , \omega b(x)\rangle$ is a reversible complement  odd length cyclic code  over  $R_{2+2\omega}$,  then both $a$ and $b$ must be self-reciprocal.
\end{thm}
\pf Let $a(x)=a_{0}+a_{1}x+\dots+a_{s}x^{s} \in C, s \leq n-1$. \\*Because $\C$ is reversible complement, we have $$a(x)^{rc}=(2+2\omega)+(2+2\omega)x+\dots+(2+2\omega)x^{n-s-2}+a_{s}^{c}x^{n-s-1}+\dots+a_{1}^{c}x^{n-2}+a_{0}^{c}x^{n-1} \in \C.$$ Using linearity of $\C$ and proposition \ref{prop:linear}, we get $$(2+2\omega)(1+x+\dots+x^{n-1})+a(x)^{rc} \in \C.$$ Then, $$[(2+2\omega)+a_{s}^{c}]x^{n-s-1}+[(2+2\omega)+a_{s-1}^{c}]x^{n-s}+\dots+[(2+2\omega)+a_{0}^{c}]x^{n-1} \in \C.$$ Since $(2+2\omega)+x^{c}=x$ for any $x\in R_{2+2\omega}$, then $$x^{n-s-1}[a_{s}+a_{s-1}x+\dots+a_{0}x^{s}]\in \C.$$ Then, $x^{n-s-1}a^{*}(x) \in \C$ which implies $a^{*}(x) \in \C$. \\*Since $\C=\langle a(x) , \omega b(x)\rangle$, then $$a^{*}(x)=a(x)c(x)+\omega b(x)d(x)$$ where $c(x) , d(x) \in Z_{4}[x]$. We have $a_{i}\in Z_{4}$ for $i\in \{0,1,\dots,s\}$, so $d(x)=0$. Then $a^{*}(x)=a(x)c(x)$. Then, $deg(a^{*}(x))= deg(a(x))$. Hence, $c(x)=m \in Z_{4}$ where $m\neq 0$. Thus, $a$ must be self-reciprocal.
\\Similarly, let $b(x)=b_{0}+b_{1}x+\dots+b_{t}x^{t} $  ,$t\leq n-1$. Then, $$\omega b(x)=\omega b_{0}+\omega b_{1}x+...+\omega b_{t}x^{t}.$$ $$(\omega b(x))^{rc}=(2+2\omega)+(2+2\omega)x+...+(2+2\omega)x^{n-t-2}+(\omega b_{t})^{c}x^{n-t-1}+\dots+(\omega b_{0})^{c}x^{n-1} \in \C.$$ Using linearity of $\C$ and Proposition \ref{prop:linear}, we get $$(\omega b(x))^{rc}-(2+2\omega)(1+x+\dots+x^{n-1})\in \C $$ Then, $$[(\omega b_{t})^{c}-(2+2\omega)]x^{n-t-1}+\dots+[(\omega b_{1})^{c}-(2+2\omega)]x^{n-2}+[(\omega b_{0})^{c}-(2+2\omega)]x^{n-1}\in \C.$$ Then, $$x^{n-t-1}[((\omega b_{t})^{c}-(2+2\omega))+\dots+((\omega b_{1})^{c}-(2+2\omega))x^{t-1}+((\omega b_{0})^{c}-(2+2\omega))x^{t}]\in \C.$$ Using Lemma \ref{lem:5}, $$x^{n-t-1}[\omega b_{t}+\dots+\omega b_{1}x^{t-1}+\omega b_{0}x^{t}]\in \C.$$ Then, $x^{n-t-1}\omega b^{*}(x)\in C$ which implies $\omega b^{*}(x)\in \C$. So, $$\omega b^{*}(x)=g(x)c{'}(x)+\omega b(x)d^{'}(x)$$ where $c^{'}(x) , d^{'}(x) \in Z_{4}[x]$. Since every term of $a(x)$ does not contain $\omega$, then $c{'}(x)=0$. Hence, $b^{*}(x)=b(x)d^{'}(x)$, which implies $deg(b^{*}(x))= deg(b(x))$ then, $d^{'}(x)=m^{'} \in Z_{4}$, where $m^{'}\neq 0$. Thus, $b(x)$ is self-reciprocal. 

\vspace*{1cm}

We found the sufficient conditions for showing the converse of Theorem \ref{thm:reverse complement}.
\begin{thm}\label{7}
If $\C=\langle a(x) , \omega b(x)\rangle$ is a cyclic code of odd length over the ring $R_{2+2\omega}$, and  if $(2+2\omega)(1+x+\dots+x^{n-1})\in \C$ and $a$ and $b$ are self reciprocal, then $\C$ must be reversible complement.
\end{thm}
\pf Let $h(x)\in \C$. We want to show that $h(x)^{rc}\in C$. \\*Let $h(x)=h_{0}+h_{1}x+...+h_{t}x^{t} \in \C   , t \leq n-1$. Since $h(x)\in \C$, then $$h(x)=a(x)c(x)+\omega b(x)d(x).$$ Then, $$h^{*}(x)=(a(x)c(x)+\omega b(x)d(x))^{*}=[a(x)c(x)]^{*}+x^{s}[\omega b(x)d(x)]^{*}$$ $$=a^{*}(x)c^{*}(x)+x^{s}[\omega b^{*}(x)d^{*}(x)].$$\\* Since $a^{*}(x)=ma(x)$ and $b^{*}(x)=m^{'}b(x)$, then $$h^{*}(x)=ma(x)c^{*}(x)+\omega b(x)[m^{'}x^{s}d^{*}(x)].$$ Hence, $h^{*}(x)\in \C$. \\*Since $\C$ is cyclic, then $$x^{n-t-1}h(x)=h_{0}x^{n-t-1}+h_{1}x^{n-t}+\dots+h_{t}x^{n-1}\in \C.$$
 \\*Since $(2+2\omega)(1+x+\dots+x^{n-1})\in \C$ and $\C$ is linear, then $$(2+2\omega)(1+x+\dots+x^{n-1})+x^{n-t-1}h(x)\in \C.$$ Then, $$(2+2\omega)+(2+2\omega)x+\dots+(2+2\omega)x^{n-1}+[h_{0}x^{n-t-1}+h_{1}x^{n-t}+\dots+h_{t}x^{n-1}]\in \C,$$ which implies $(2+2\omega)+(2+2\omega)x+\dots+(2+2\omega)x^{n-t-2}+[(2+2\omega)+h_{0}]x^{n-t-1}+[(2+2\omega)+h_{1}]x^{n-t}+\dots+[(2+2\omega)+h_{t}]x^{n-1}\in \C.$ Since, $(2+2\omega )+x=x^{c}$, then $$(2+2\omega)+(2+2\omega)x+\dots+(2+2\omega)x^{n-t-2}+h_{0}^{c}x^{n-t-1}+h_{1}^{c}x^{n-t}+\dots +h_{t}^{c}x^{n-1}\in \C.$$ Hence, $(h^{*}(x))^{rc}\in \C$. We have $h^{*}(x)\in \C$ for any polynomial $h(x)\in \C$. Then, $$((h^{*}(x))^{rc})^{*}\in \C.$$ $$((h^{*}(x))^{rc})^{*}=h_{t}^{c}+h_{t-1}^{c}x+\dots+h_{0}^{c}x^{t}+(2+2\omega)x^{t+1}+\dots+(2+2\omega)x^{n-1}\in \C.$$ Note that $((h^{*}(x))^{rc})^{*}=(x^{n-t-1}h(x))^{rc}\in \C$ $\implies$ $h(x)^{rc}\in \C$. Thus, $\C$ is reversible complement.

\vspace*{1cm}

Using \cite[ Lemmas 2 and 3]{1}, we obtain the following:

\begin{lem} \label{lem:cy1}
A necessary and  sufficient condition that the DNA cyclic code $\C$ over $R_{2+2\omega}$ be reversible is that $x^{s}\phi^{-1}(\phi(g(x)^{r}))\in \C$ for all $s$ where $s\in\{0, 1,\dots, k-1\}$ and $g(x)$ is the first row in the generator matrix of $\C$.
\end{lem}

\begin{lem} \label{lem:cy2}
The DNA cyclic code $\C$ over $R_{2+2\omega}$ is complementary if $(2+2\omega)(1+x+\dots+x^{n-1})\in \C$ and conversely.
\end{lem}

Combining Lemmas \ref{lem:cy1} and \ref{lem:cy2} we obtain the following;

\begin{prop} \label{prop:cy3}
A necessary  and sufficient condition that the DNA cyclic code $\C$ over $R_{2+2\omega}$ be reversible complement is that $x^{s}\phi^{-1}(\phi(g(x)^{r}))\in \C$ and $(2+2\omega)(1+x+\dots+x^{n-1})\in \C$
\end{prop}

 \section{Quasi-Cyclic Codes over $R_{2+2\omega}$}
 \label{sec:3}
 Every $l$-quasi-cyclic code has a generator matrix

$$G=\begin{bmatrix}
g(x) \\
x^{l}g(x) \\
x^{2l}g(x)\\
.\\
.\\
.\\
x^{(k-1)l}g(x)
\end{bmatrix}$$

A quasi-cyclic code has length $n=l\cdot m$ ,where $l$ is the index and $m$ is the co-index, and the number of rows $k\leq m$.

\vspace*{1cm}

Similarly, using \cite[Lemmas 2 and 3 ]{1}, we obtain the following:

\begin{lem}\label{lem:quasi}
A necessary and sufficient condition that the DNA quasi-cyclic code over $R_{2+2\omega}$ be reversible is  that  $x^{s l}\phi^{-1}(\phi(g(x)^{r}))\in Q$ for all $s$, where $s\in\{0, 1,\dots, k-1\}$ and $g(x)$ is any row in the generator matrix of $Q$.
\end{lem}

\begin{lem}\label{lem:quasi1}
The DNA quasi-cyclic code $Q$ over $R_{2+2\omega}$ is complement if and only if $(2+2\omega)(1+x+\dots+x^{n-1})\in Q$.
\end{lem}

From Lemmas \ref{lem:quasi} and \ref{lem:quasi1} this holds:
\begin{prop}
A DNA code  which is quasi-cyclic over  $R_{2+2\omega}$ has  all three reversible complement, reversible and complement conditions if and only if $x^{s l}\phi^{-1}(\phi(g(x)^{r}))\in Q$ and $(2+2\omega)(1+x+\dots+x^{n-1})\in Q$.
\end{prop}

The following conjecture presents sufficient conditions to obtain reversible and reversible complement DNA quasi-cyclic codes.
\begin{conjecture} \label{con:conj}
A DNA code which is quasi-cyclic with length $2n=2(l\cdot m)$ where $l>1$ is reversible, also reversible complement if and only if $l\lneq m$ and there exists a row $g(x)=\sum_{i=0}^{n-1}g_{i}x^{i}$ in the generator matrix $G$ such that $g_{i}=g_{n-i-1}$ for all $i\in \{0,1,\dots,n-1\}$.
\end{conjecture}

Note that in the above conjecture, it is sufficient to know any matrix that generates  the quasi-cyclic code to determine whether the corresponding DNA code meets or does not meet the combinatorial constraints.

\begin{exam} 

We can get the 2-quasi-cyclic DNA code with parameters $(2n=16, M=256, d_{H}=8)$ by  shifting the last two coordinates of the vector $(0, 3, 1, 2, 2, 1, 3, 0)$. These 256 DNA codewordes are as follows:
\begin{scriptsize}
\[ \begin{array}{lcrr}
\mbox{$AAAAAAAAAAAAAAAA$} & \mbox{$AGGGGGAGGAAAAAGA$} & \mbox{$GGAAAAGGGGAAAAGG$} & \mbox{$CTAAAACTTCGGGGTC$}\\
\mbox{$GAGGGGGAAGAAAAAG$} & \mbox{$TGCCCCTGGTAAAAGT$} & \mbox{$TATTTTTAATAAAAAT$}& \mbox{$TTGGGGTTTTGGGGTT$}\\
\mbox{$CACCCCCAACAAAAAC$} & \mbox{$CGTTTTCGGCAAAAGC$} & \mbox{$CCAAAACCCCAAAACC$} & \mbox{$TCAAAATCCTGGGGCT$}\\
\mbox{$CTGGGGCTTCAAAATC$} & \mbox{$TTAAAATTTTAAAATT$} & \mbox{$TCGGGGTCCTAAAACT$} & \mbox{$GTTTTTGTTGGGGGTG$}\\
\mbox{$GTCCCCGTTGAAAATG$} & \mbox{$GCTTTTGCCGAAAACG$} & \mbox{$ATTTTTATTAAAAATA$} & \mbox{$GCCCCCGCCGGGGGCG$}\\
\mbox{$GGAGGAAAAAGAAGGG$} & \mbox{$GAGAAGAGGAGAAGAG$} & \mbox{$AAAGGAGGGGGAAGAA $} & \mbox{$ACTTTTACCAGGGGCA$}\\
\mbox{$AGGAAGGAAGGAAGGA$} & \mbox{$CACTTCTGGTGAAGAC$} & \mbox{$CGTCCTTAATGAAGGC$} & \mbox{$ATCCCCATTAGGGGTA$}\\
\mbox{$TGCTTCCAACGAAGGT$} & \mbox{$TATCCTCGGCGAAGAT$} & \mbox{$TTAGGACCCCGAAGTT$} &\mbox{$GGGAAGAAAAAGGAGG$}\\
\mbox{$TCGAAGCTTCGAAGCT$} & \mbox{$GAAAAAGAAGGGGGAG $} & \mbox{$TGTTTTTGGTGGGGGT $} & \mbox{$GAAGGAAGGAAGGAAG$}\\
\mbox{$CTGAAGTCCTGAAGTC$} & \mbox{$ACCTTCGTTGGAAGCA$} & \mbox{$ATTCCTGCCGGAAGTA$} & \mbox{$AAGAAGGGGGAGGAAA$}\\
\mbox{$GTCTTCACCAGAAGTG$} & \mbox{$GCTCCTATTAGAAGCG$} & \mbox{$AAGGGGAAAAGGGGAA$} & \mbox{$AGAGGAGAAGAGGAGA$}\\
\mbox{$AGAAAAAGGAGGGGGA$} & \mbox{$GGGGGGGGGGGGGGGG$} & \mbox{$CCGGGGCCCCGGGGCC $} & \mbox{$CATCCTTGGTAGGAAC$}\\
\mbox{$TACCCCTAATGGGGAT $} & \mbox{$CATTTTCAACGGGGAC $} & \mbox{$CGCCCCCGGCGGGGGC $} &\mbox{$CGCTTCTAATAGGAGC$}\\
\mbox{$TACTTCCGGCAGGAAT$} & \mbox{$TTGAAGCCCCAGGATT $} & \mbox{$TCAGGACTTCAGGACT $} & \mbox{$TGTCCTCAACAGGAGT$}\\
\mbox{$CCGAAGTTTTAGGACC $} & \mbox{$CTAGGATCCTAGGATC $} & \mbox{$ACTCCTGTTGAGGACA $} & \mbox{$ATCTTCGCCGAGGATA $}\\
\mbox{$GTTCCTACCAAGGATG $} & \mbox{$GCCTTCATTAAGGACG $} & \mbox{$CCTGGTAAAAGTTGCC  $} & \mbox{$CTCAACAGGAGTTGTC $}\\
\mbox{$TTTGGTGGGGGTTGTT $} & \mbox{$TCCAACGAAGGTTGCT $} & \mbox{$GTGTTGTGGTGTTGTG $} & \mbox{$GCACCATAATGTTGCG $}\\
\mbox{$ACGTTGCAACGTTGCA $} & \mbox{$ATACCACGGCGTTGTA $} & \mbox{$AATGGTCCCCGTTGAA $} &\mbox{$AGCAACCTTCGTTGGA $} \\
\mbox{$GGTGGTTTTTGTTGGG $} & \mbox{$GACAACTCCTGTTGAG $} & \mbox{$TGGTTGGTTGGTTGGT $} & \mbox{$TAACCAGCCGGTTGAT $}\\
\mbox{$CAGTTGACCAGTTGAC $} & \mbox{$CGACCAATTAGTTGGC $} & \mbox{$TTTAATAAAAATTATT $} & \mbox{$TCCGGCAGGAATTACT $}\\
\mbox{$CCTAATGGGGATTACC $} & \mbox{$CTCGGCGAAGATTATC $} & \mbox{$ACGCCGTGGTATTACA  $} & \mbox{$ATATTATAATATTATA $}\\
\mbox{$GTGCCGCAACATTATG $} & \mbox{$GCATTACGGCATTACG $} & \mbox{$GGTAATCCCCATTAGG $} & \mbox{$GACGGCCTTCATTAAG $}\\
\mbox{$AATAATTTTTATTAAA$} & \mbox{$AGCGGCTCCTATTAGA$} & \mbox{$CAGCCGGTTGATTAAC$} & \mbox{$CGATTAGCCGATTAGC$}\\
\mbox{$TGGCCGACCAATTAGT$} & \mbox{$TAATTAATTAATTAAT$} & \mbox{$CCCAACAAAAACCACC$} & \mbox{$CTTGGTAGGAACCATC$}\\
\mbox{$TTCAACGGGGACCATT$} & \mbox{$TCTGGTGAAGACCACT$} & \mbox{$GTACCATGGTACCATG$} & \mbox{$GCGTTGTAATACCACG$}\\
\mbox{$ACACCACAACACCACA$} & \mbox{$ATGTTGCGGCACCATA$} & \mbox{$AACAACCCCCACCAAA$} & \mbox{$AGTGGTCTTCACCAGA$}\\
\mbox{$GGCAACTTTTACCAGG$} & \mbox{$GATGGTTCCTACCAAG $} & \mbox{$TGACCAGTTGACCAGT $} & \mbox{$TAGTTGGCCGACCAAT $}\\
\mbox{$CAACCAACCAACCAAC $} & \mbox{$CGGTTGATTAACCAGC $} & \mbox{$TTCGGCAAAAGCCGTT $} & \mbox{$TCTAATAGGAGCCGCT $}\\
\mbox{$CCCGGCGGGGGCCGCC $} & \mbox{$CTTAATGAAGGCCGTC $} & \mbox{$ACATTATGGTGCCGCA $} & \mbox{$ATGCCGTAATGCCGTA $}\\
\mbox{$GTATTACAACGCCGTG $} & \mbox{$GCGCCGCGGCGCCGCG $} & \mbox{$GGCGGCCCCCGCCGGG $} & \mbox{$GATAATCTTCGCCGAG $}\\
\mbox{$AACGGCTTTTGCCGAA $} & \mbox{$AGTAATTCCTGCCGGA $} & \mbox{$CAATTAGTTGGCCGAC  $} & \mbox{$CGGCCGGCCGGCCGGC $}\\
\mbox{$TGATTAACCAGCCGGT $} & \mbox{$TAGCCGATTAGCCGAT $} & \mbox{$AACCCCAAAACCCCAA $} & \mbox{$AGTTTTAGGACCCCGA $}\\
\mbox{$GGCCCCGGGGCCCCGG $} & \mbox{$GATTTTGAAGCCCCAG $} & \mbox{$TGAAAATGGTCCCCGT $} & \mbox{$TAGGGGTAATCCCCAT $}\\
\mbox{$CAAAAACAACCCCCAC $} & \mbox{$CGGGGGCGGCCCCCGC $} & \mbox{$CCCCCCCCCCCCCCCC $} & \mbox{$CTTTTTCTTCCCCCTC $}\\
\mbox{$TTCCCCTTTTCCCCTT $} & \mbox{$TCTTTTTCCTCCCCCT $} & \mbox{$GTAAAAGTTGCCCCTG  $} & \mbox{$GCGGGGGCCGCCCCCG $}\\
\mbox{$ACAAAAACCACCCCCA $} & \mbox{$ATGGGGATTACCCCTA $} & \mbox{$GGCTTCAAAATCCTGG $} & \mbox{$GATCCTAGGATCCTAG $}\\
\mbox{$AACTTCGGGGTCCTAA $} & \mbox{$AGTCCTGAAGTCCTGA $} & \mbox{$CAAGGATGGTTCCTAC $} & \mbox{$CGGAAGTAATTCCTGC $}\\
\mbox{$TGAGGACAACTCCTGT $} & \mbox{$TAGAAGCGGCTCCTAT $} & \mbox{$TTCTTCCCCCTCCTTT $} & \mbox{$TCTCCTCTTCTCCTCT $}\\
\mbox{$CCCTTCTTTTTCCTCC$} & \mbox{$CTTCCTTCCTTCCTTC$} & \mbox{$ACAGGAGTTGTCCTCA$} & \mbox{$ATGAAGGCCGTCCTTA$}\\
\mbox{$GTAGGAACCATCCTTG$} & \mbox{$GCGAAGATTATCCTCG$} & \mbox{$AATTTTAAAATTTTAA$} & \mbox{$AGCCCCAGGATTTTGA$}\\
\mbox{$GGTTTTGGGGTTTTGG$} & \mbox{$GACCCCGAAGTTTTAG$} & \mbox{$TGGGGGTGGTTTTTGT$} & \mbox{$TAAAAATAATTTTTAT$}\\
\mbox{$CAGGGGCAACTTTTAC$} & \mbox{$CGAAAACGGCTTTTGC$} & \mbox{$CCTTTTCCCCTTTTCC$} & \mbox{$CTCCCCCTTCTTTTTC$}\\
\mbox{$TTTTTTTTTTTTTTTT $} & \mbox{$TCCCCCTCCTTTTTCT $} & \mbox{$GTGGGGGTTGTTTTTG $} & \mbox{$GCAAAAGCCGTTTTCG $}\\
\mbox{$ACGGGGACCATTTTCA $} & \mbox{$ATAAAAATTATTTTTA $} & \mbox{$GGTCCTAAAACTTCGG  $} & \mbox{$GACTTCAGGACTTCAG$}\\
\mbox{$AATCCTGGGGCTTCAA$} & \mbox{$AGCTTCGAAGCTTCGA $} & \mbox{$CAGAAGTGGTCTTCAC $} & \mbox{$CGAGGATAATCTTCGC $}\\
\mbox{$TGGAAGCAACCTTCGT $} & \mbox{$TAAGGACGGCCTTCAT $} & \mbox{$TTTCCTCCCCCTTCTT $} & \mbox{$TCCTTCCTTCCTTCCT $}\\
\mbox{$CCTCCTTTTTCTTCCC $} & \mbox{$CTCTTCTCCTCTTCTC $} & \mbox{$ACGAAGGTTGCTTCCA $} & \mbox{$ATAGGAGCCGCTTCTA $}\\
\mbox{$GTGAAGACCACTTCTG $} & \mbox{$GCAGGAATTACTTCCG $} & \mbox{$CCGTTGAAAATGGTCC $} & \mbox{$CTACCAAGGATGGTTC $}\\
\mbox{$TTGTTGGGGGTGGTTT $} & \mbox{$TCACCAGAAGTGGTCT $} & \mbox{$GTTGGTTGGTTGGTTG  $} & \mbox{$GCCAACTAATTGGTCG $}\\
\mbox{$ACTGGTCAACTGGTCA$} & \mbox{$ATCAACCGGCTGGTTA$} & \mbox{$AAGTTGCCCCTGGTAA$} & \mbox{$AGACCACTTCTGGTGA $}\\
\mbox{$CCAGGATTTTGAAGCC $} & \mbox{$GGGTTGTTTTTGGTGG$} & \mbox{$GAACCATCCTTGGTAG$} & \mbox{$TGTGGTGTTGTGGTGT$}\\
\mbox{$TACAACGCCGTGGTAT$} & \mbox{$CATGGTACCATGGTAC$} & \mbox{$CGCAACATTATGGTGC$} & \mbox{$TTGCCGAAAACGGCTT$}\\
\mbox{$TCATTAAGGACGGCCT $} & \mbox{$CCGCCGGGGGCGGCCC $} & \mbox{$CTATTAGAAGCGGCTC $} & \mbox{$ACTAATTGGTCGGCCA $}\\
\mbox{$ATCGGCTAATCGGCTA $} & \mbox{$GTTAATCAACCGGCTG $} & \mbox{$GCCGGCCGGCCGGCCG  $} & \mbox{$GGGCCGCCCCCGGCGG $}\\
\mbox{$GAATTACTTCCGGCAG $} & \mbox{$AAGCCGTTTTCGGCAA $} & \mbox{$AGATTATCCTCGGCGA $} & \mbox{$CATAATGTTGCGGCAC $}\\
\mbox{$CGCGGCGCCGCGGCGC $} & \mbox{$TGTAATACCACGGCGT $} & \mbox{$TACGGCATTACGGCAT $} & \mbox{$CCACCAAAAACAACCC $}\\
\mbox{$CTGTTGAGGACAACTC $} & \mbox{$TTACCAGGGGCAACTT $} & \mbox{$TCGTTGGAAGCAACCT $} & \mbox{$GTCAACTGGTCAACTG $}\\
\mbox{$GCTGGTTAATCAACCG $} & \mbox{$ACCAACCAACCAACCA $} & \mbox{$ATTGGTCGGCCAACTA $} & \mbox{$AAACCACCCCCAACAA $}\\
\mbox{$AGGTTGCTTCCAACGA $} & \mbox{$GGACCATTTTCAACGG $} & \mbox{$GAGTTGTCCTCAACAG  $} & \mbox{$TGCAACGTTGCAACGT$}\\
\mbox{$TATGGTGCCGCAACAT$} & \mbox{$CACAACACCACAACAC$} & \mbox{$CGTGGTATTACAACGC$} & \mbox{$TTATTAAAAATAATTT$}\\
\mbox{$TCGCCGAGGATAATCT $} & \mbox{$CCATTAGGGGTAATCC $} & \mbox{$CTGCCGGAAGTAATTC $} & \mbox{$ACCGGCTGGTTAATCA $}\\
\mbox{$ATTAATTAATTAATTA $} & \mbox{$GTCGGCCAACTAATTG $} & \mbox{$GCTAATCGGCTAATCG $} & \mbox{$GGATTACCCCTAATGG $}\\
\mbox{$GAGCCGCTTCTAATAG $} & \mbox{$AAATTATTTTTAATAA $} & \mbox{$AGGCCGTCCTTAATGA $} & \mbox{$CACGGCGTTGTAATAC $}\\
\mbox{$CGTAATGCCGTAATGC $} & \mbox{$TGCGGCACCATAATGT $} & \mbox{$ACCCCCACCAAAAACA $} & \mbox{$TATAATATTATAATAT $}
\end{array}\] 
\end{scriptsize}
\end{exam}
In similar ways, we have found many examples of quasi-cyclic DNA codes of indices $2$ and $3$ which are reversible and reversible complement given in table \ref{tab:3} and table \ref{tab:4}. For the quasi-cyclic codes of index $3$, we will shift the last $3$ coordinates of the vector.
newpage
\begin{table}
  \caption{Quasi-Cyclic DNA Code of index $2$}
\label{tab:3}       
\begin{tabular}{ll}
\hline\noalign{\smallskip}
First row vector & $(2n, M, d_{H})$  \\
\noalign{\smallskip}\hline\noalign{\smallskip}
$(1, 2, 2, 1)$ & $(8, 256, 2)$   \\
$(1, 1, 2\omega+3, 2, 2, 2\omega+3)$ & $(12, 1024, 4)$  \\
$(\omega, 1, 1, \omega, 3, 3)$ & $(12, 2048, 4)$ \\
$(0, 3, 1, 2, 2, 1, 3, 0)$ & $(16, 256, 8)$ \\
\noalign{\smallskip}\hline
\end{tabular}
\end{table}

\begin{table}
  \caption{Quasi-Cyclic DNA Code of index $3$}
\label{tab:4}       
\begin{tabular}{ll}
\hline\noalign{\smallskip}
First row vector & $(2n, M, d_{H})$  \\
\noalign{\smallskip}\hline\noalign{\smallskip}
$(3, 3, 3, 1, 1, 1, 1, 1, 1)$ & $(18, 256, 6)$   \\
$(3+\omega, 3+\omega, 3+\omega, 1, 1, 1, 1, 1, 1, 3+\omega, 3+\omega, 3+\omega,)$ & $(24, 512, 6)$  \\
\noalign{\smallskip}\hline
\end{tabular}
\end{table}
 Using the Sphere Packing Bound defined in \cite{3}, we found that Table 3 and Table 4 contain good codes.

\section{General Results}
 \label{sec:4}
 The structure of the sixteen rings $R_\theta=\Z_{4}+\omega \Z_{4} \  (\omega^{2}=\theta)$ where $\theta \in \Z_{4}+\omega \Z_{4}$ has been studied in \cite{3}. We found in addition, the following general properties:
\begin{itemize}
\item[-] There are special zero divisors which are self-invertible under addition; $0, 2, 2\omega$ and $2+2\omega$ will be the same in any ring $R_{\theta}$. 
\item[-] There are special units, $1, 3, 1+2\omega$ and $3+2\omega$, which are self-invertible under multiplication, that  behave the same for any ring $R_{\theta}$.
\item[-] The ring $R_{\theta}$ is a Frobenius ring for $\theta \in \{0, 1, 2, 3, \omega, 1+\omega, 2+\omega, 3+\omega, 2\omega, 1+2\omega, 2+2\omega,3\omega, 1+3\omega, 2+3\omega, 3+3\omega \} $.

\end{itemize}
 We use the generalized Gau map which has been defined in \cite{3} to generalize some of the results to the ring $R_\theta$.

\begin{prop}
For any reversible complement linear code $\CC$ over the ring $R_\theta$, the string $\boldsymbol{\lambda}$$\in \CC$  where $\boldsymbol{\lambda}$ is a string with each coordinte $\lambda$  where $\lambda \in \{2, 2\omega, 2+2\omega\}$.
\end{prop}

\begin{thm}
Suppose that $\C$ is cyclic in $R_{\theta}[x]/<x^{n}-1>$. Then there exists polynomials $a,  b$ in $R_{\theta}[x]/<x^{n}-1>$ such that $b(x)|a(x)|x^{n}-1$ and $\C=<a(x), \omega b(x)>$.
\end{thm}

\begin{lem}
Let $a$ , $b$ be two polynomials in $R_{\theta}[x]$ with the property that $deg(a(x))-deg(b(x))=t$, then:
\item[-]$(a(x)b(x))^{*}=a^{*}(x)b^{*}(x)$.
\item[-]$(a(x)+b(x))^{*}=a^{*}(x)+x^{t}b^{*}(x)$.
\end{lem}

\begin{thm}\label{4}
Let $\C=\langle a(x) , \omega b(x)\rangle$ be a cyclic code over the ring $R_\theta $.  It follows that the necessary and sufficient condition for $\C$ to be reversible is that both $a$ and $b$ be self-reciprocal.
\end{thm}

\begin{lem}
For any $x\in R_{\theta} ; \ \omega x=(\omega x)^{c}-(\lambda)$ where $\lambda \in \{2, 2\omega, 2+2\omega\}$.
\end{lem}

\begin{thm}\label{5}
If $\C=\langle a(x) , \omega b(x)\rangle$ is a reversible complement  cyclic code over $R_\theta$ having odd  length,  then both $a$ and $b$ are self-reciprocal.
\end{thm}

\begin{thm}\label{6}
Let $\C=\langle a(x) , \omega b(x)\rangle$ be a cyclic code of odd length over the ring $R_\theta$. If $\lambda(1+x+...+x^{n-1})\in \C$ where $\lambda \in  \{2, 2\omega , 2+2\omega\}$ and $a(x)$ and $b(x)$ are self reciprocal, then $\C$ is reversible complement.
\end{thm}

Using \cite[Lemma 10 and Remark 12]{3}, we obtain:
\begin{lem} \label{Cy1}
A necessary and sufficient condition that a DNA cyclic code $\C$ over $R_{\theta}$ be reversible is that $x^{s}\phi^{-1}(\phi(g(x)^{r}))$ $\in \C$ for all $s$, where $s\in\{0, 1,\dots, k-1\}$ and $g(x)$ is the first row in a generator matrix of $\C$.
\end{lem}

\begin{lem} \label{Cy2}
The DNA cyclic code $\C$ over $R_{\theta}$ is complement if and only if $\lambda(1+x+...+x^{n-1})\in \C$ where $\lambda \in \{2, 2\omega, 2+2\omega\}$.
\end{lem}

Combining Lemmas \ref{Cy1} and \ref{Cy2}, we obtain the following;
\begin{prop}
The DNA cyclic code $\C$ over the ring $R_{\theta}$ is reversible complement, reversible and complement if and only if $x^{s}\phi^{-1}(\phi(g(x)^{r}))\in \C$ and $\lambda(1+x+\dots+x^{n-1})\in \C$ where $\lambda \in \{2, 2\omega, 2+2\omega\}$.
\end{prop}

Similarly, using \cite[Lemma 10 and Remark 12]{3}, we obtain the following:
\begin{lem} \label{Q1}
The DNA quasi-cyclic code $Q$ over $R_{\theta}$ is reversible if and only if  \newline
$x^{s l}\phi^{-1}(\phi(g(x)^{r}))\in Q$ for all $s$, where $s\in\{0, 1,\dots,k-1\}$ and $g(x)$ is any row in the generator matrix of $Q$.
\end{lem}

\begin{lem} \label{Q2}
The DNA quasi-cyclic code $Q$ over $R_{\theta}$ is complement if and only if \newline
$\lambda(1+x+...+x^{n-1})\in Q$ where $\lambda \in \{2, 2\omega, 2+2\omega\}$.
\end{lem}

Combining Lemmas \ref{Q1} and \ref{Q2} we obtain:
\begin{prop}
The DNA quasi-cyclic code $Q$ over the ring $R_{\theta}$ is reversible complement, reversible and complement if and only if $x^{s l}\phi^{-1}(\phi(g(x)^{r}))\in Q$ and $\lambda(1+x+\dots+x^{n-1})\in Q$ where $\lambda \in \{2, 2\omega, 2+2\omega\}$.
\end{prop}

Self-dual codes over the ring $R_{i}=\Z_{4}+\omega \Z_{4}$, where $\omega^{2}=i$ for $i\in \{1, 2\omega\}$, have been studied in \cite{4}. We generalize some results obtained in that paper using the Gray map and the generalized Gau map.

\vspace*{0.25cm}
In this paper, we consider the usual inner product regarding to the duality concepts.
\begin{definition} 
The dual code $\CC^{\perp}$ of a code $\CC$ over a ring $R$ is defined as $\CC^{\perp}=\{c\in R^{n}\mid c\cdot c'=0, \forall c'\in \CC \}$, where $c\cdot c'$ denotes the usual inner product .
\end{definition}

\begin{definition}
A code $\CC$ over a ring is called self-dual  if $\CC=\CC^{\perp}$.
\end{definition}

The Gray map $\psi_{i}$ is defined in \cite{4} for the ring $R_{i}$ where $i\in \{1, 2\omega\}$. We extend the definition of Gray map for the ring $R_{\theta}$, so the Gray map $\psi_{\theta} : R_{\theta}\rightarrow \Z_{4}^{2}$ is defined as: $\psi_{\theta}(a+\omega b)=(b, a+b)$. This is extended component wise to $\psi_{\theta} : R_{\theta}^{n}\rightarrow \Z_{4}^{2n}$.

\begin{thm} Let $\CC$ be a linear code over the ring $R_{j}$ where $j\in \{2, 2\omega\}$, Then $\psi_{j}(\CC^{\perp})=\psi_{j}(\CC)^{\perp}$.
\end{thm}
\pf The proof of this theorem for $j=2\omega$ is given in \cite[Theorem 2.1]{4}. Consider $j=2$ and assume that $\phi(c) \in \phi(\CC^{\perp})$, where $c=a+\omega b=(a_{0}, a_{1},\dots ,a_{n-1})+\omega (b_{0}, b_{1},\dots ,b_{n-1}) \in \CC^{\perp}$, it follows that $c\cdot c'=0$ for all $c'=a'+\omega b'\in \CC$ which implies $a\cdot a'+2(b\cdot b')=0$ and $a\cdot b'+b\cdot a'=0$. Then $\psi_{2}(c)\cdot\psi_{2}(c')=(b, a+b)\cdot(b', a'+b')=a\cdot a'+2(b\cdot b')+(a\cdot b'+a'\cdot b)=0$. Therefore, $\psi_{2}(\CC^{\perp})\subseteq \psi_{2}(\CC)^{\perp}$.\\* The second part of the proof is by similar technique as proving \cite[Theorem 2.1]{4}.

\begin{cor}
Whenever $\CC$ is self-dual code over the ring $R_{j}$ where $j\in \{2, 2\omega\}$, Then $\psi_{j}(\CC)$ must be also self-dual over $\Z_{4}$
\end{cor}
\pf Let $\CC=\CC^{\perp}$, then $\psi_{j}(\CC)=\psi_{j}(\CC^{\perp})$ and $\psi_{j}(\CC^{\perp})=\phi_{j}(\CC)^{\perp}$.

\vspace*{0.5cm}
Now we consider the generalized Gau map $\phi$ for the following results.
\begin{definition} \label{def:dual of DNA code}
The dual code of a DNA code $\phi(\CC)$ where $\CC$ is a code over $R_{\theta}$ is defined as  $\phi(\CC)^{\perp}=\{\phi(c)\mid c\cdot c'=0, \forall c'\in \CC \}$.
\end{definition}

We define the self-dual DNA code for a DNA strings of even length
\begin{definition} A DNA code $\phi(\CC)$ is self-dual if $\phi(\CC)=\phi(\CC)^{\perp}$
\end{definition}

one can easily show the following remark which is clear by definition \ref{def:dual of DNA code}
\begin{rem}\label{rem:7}
$\phi(\CC)^{\perp}=\phi(\CC^{\perp})$ for all linear codes $\CC$ over the ring $R_{\theta}$.
\end{rem}

\begin{cor}
$\CC$ is a self-dual code over the ring $R_{\theta}$ if and only if $\phi(\CC)$ is self-dual.
\end{cor}
\pf $\CC=\CC^{\perp}\Leftrightarrow \phi(\CC)=\phi(\CC^{\perp})=\phi(\CC)^{\perp}$.

\begin{definition}
A self-dual code that is free is just one with no non-zero codeword made up of only zero divisors.
\end{definition}

\begin{exam} Let $\CC$ be a linear code over the ring $R_{2}$ generated by 
$$G=\begin{bmatrix}
1&0 & 0& 0& 1& 1+2\omega &1+2\omega &0 \\
0&1& 0&0 &3+2\omega & 1& 0&1+2\omega \\
0&0&1&0&1+2\omega&0&3& 1+2\omega\\
0&0&0&1&0&3+2\omega&1+2\omega&1
\end{bmatrix}$$
This is an $(8,16^{4})-$ free self-dual code over $R_{2}$.
\end{exam}

\begin{thm}\label{8}
$\frac{n}{2}+1$  is the largest possible value of minimum Hamming distance of a free self-dual code of length $n$ over $R_\theta $. 
\end{thm}
\pf The proof has the same technique \cite[Theorem 3.8]{4}.
\vspace*{1cm}
\\*We have also found the upper bound of the Gau distance $d_{G(\theta)}$ of a free self-dual code over $R_\theta $ using the following relation between the Gau and Hamming distances.
\begin{lem} \label{relation}
$d_{G(\theta)}(x,y)=d_H(x,y)+r$ for all $x,y\in \CC$, where $\CC$ is a code over the ring $R_{\theta}$ and $r$ is how many coordinates $i$ with $d_{G(\theta)}(x_{i}, y_{i})=2$, where $i\in \{0, \dots, n-1\}$.
\end{lem}
\pf Let $x,y$ be arbitrary codewords in $\CC$ of length $n$,
\\* $x=x_{0}\dots x_{n-1}$ and $y=y_0\dots y_{n-1}$.
\\Case $1$: If $d_{G(\theta)}(x_{i}, y_{i})\neq 2$ for all $i\in \{0, 1, \dots n-1\}$, \\* then, $r=0$ and $d_H(x_{i}, y_{i})=d_{G(\theta)}(x_{i}, y_{i})$ for all $i\in \{0, 1, \dots  n-1\}$.$$ d_{G(\theta)}(x,y)=\sum_{i=0}^{ n-1} d_{G(\theta)}(x_{i}, y_{i})=\sum_{i=0}^{ n-1} d_H(x_i,y_i). $$ \\*Thus, $d_{G(\theta)}(x,y)=d_H(x,y)+r$.
\\Case $2$:  If there exists $i\in\{0,1,\dots,n-1\}$ such that $d_{G(\theta)}(x_{i}, y_{i})=2$.
\\*Let $$S=\{i: d_{G(\theta)}(x_{i}, y_{i})\neq2\}$$ and let $$T=\{j: d_{G(\theta)}(x_{j}, y_{j})=2\}.$$ \\*Hence, $d_H(x_i,y_i)=d_{G(\theta)}(x_{i}, y_{i})$ for all $ i \in S$ and $d_H(x_{j}, y_{j})=1$ for all $j \in T$.
$$d_{G(\theta)}(x, y)=\sum_{i\in S} d_{G(\theta)}(x_{i}, y_{i})+\sum_{j\in T} d_{G(\theta)}(x_{j}, y_{j}),~~........~(*)$$
where $$\sum_{i\in S} d_{G(\theta)}(x_{i}, y_{i})=\sum_{i\in S} d_H(x_{i}, y_{i}),$$
and $$\sum_{j\in T} d_{G(\theta)}(x_{j}, y_{j})=2\sum_{j\in T} d_H(x_{j}, y_{j})=\sum_{j\in T} d_H(x_{j}, y_{j})+r.$$ 
By substituting in $(*)$, we obtain  
$$d_{G(\theta)}(x, y)=\sum_{i\in S} d_H(x_{i}, y_{i})+\sum_{j\in T} d_H(x_{j}, y_{j})+r.$$ Thus, $d_{G(\theta)}(x, y)=d_H(x, y)+r$.

\begin{thm} $\frac{n}{2}+r+1$ is an upper bound for the Gau distance of a free self-dual code $\CC$ over  $R_\theta $.
\end{thm}
\pf If $\CC$ is a code over $R_\theta$ and $x$ and $y$ are arbitrary codewords in $\CC$, then by Theorem \ref{8}, $d_H(x,y)\leq\frac{n}{2}+1 $ and by Lemma \ref{relation}, \\*$d_{G(\theta)}(x,y)=d_H(x,y)+r \leq\frac{n}{2}+1+r $. Thus, $d_{G(\theta)}(x,y)\leq \frac{n}{2}+r+1$.

\section{Conclusion}
 \label{sec:5}
This paper investigated cyclic and quasi-cyclic DNA codes. Conditions for DNA codes that are cyclic or quasi-cyclic have having certain combinatorial constraints have been explored. Many examples of good DNA quasi-cyclic codes have been provided.  it would be nice to study twisted and quasi-twisted DNA codes over $R_{\theta}$. Moreover, one can ascertain whether Conjecture \ref{con:conj} is true.


%
%


%
%



\end{document}